# HSBC 1950 to 2025: Conquering the world from British Hong Kong and London

by Christopher Mantzaris[*], Prof. Dr Ajda Fošner[*].

**Abstract** & introduction:
The Hongkong and Shanghai Banking Co (HSBC) just survived a civil war intermitted by World War II. By the 1950s, it obviously needed to close all its branches in Mao's People's Republic of China, yet could somehow hold its Shanghai branch, which continued likely in the shadows, as non-state banking was illegalised and even simple land owners were executed merely for being labelled "capitalist". This Asia-focused bank –in spite of it all– grew from these conditions into the behemoth it is today. Part of the growth was based on the economic boom in its core market Hong Kong, to which HSBC likely also contributed. To expand and diversify, HSBC continued the growth strategy that already started since its early days in the 1860s, this time just also inorganically: It acquired other banks, in most cases fully and in other regions. The most important acquisition was the takeover of the roughly equally-sized UK-based Midland Bank; for the following reasons: 1) It came just a year after the 1991 change of HSBC's headquarters and place of incorporation to London, so HSBC could smoothly integrate with Midland. This step also came with an additional listing of securities in London, providing HSBC funds. 2) These funds were used efficiently without much idling for the humongous acquisition. 3) The preceding decade of Margaret Thatcher's banking and finance deregulation in the UK created a beneficial environment for HSBC. 4) HSBC was proven right by the developments in Hong Kong, where the Communist Party of China illegally eroded democracy, the rule of law and civil liberties since 2020, despite promising to maintain these at least until 1 July 2047. A list of likely reasons for HSBC's prosperity and stability in face of the at times hostile environments is also provided, from which business lessons can be drawn.

## Cold war and the world globalises once more: HSBC 1950-1990s

The closing of branches in the People's Republic of China that started in 1949 was completed in 1955, by which only Shanghai remained as HSBC's sole branch there.[1] In light of the reign of Mao Zedong it as astonishing that even a single branch could be maintained.

Mao reigned from 1949 to his death in 1976 as a brutal dictator, as Mao and his policies killed up to 80 Million people in China, almost all of whom had the same citizenship as him.[2] On one hand, many of these deaths were caused by simply bad –some would say asinine– policy: At least 45 million died during Mao's *Great Leap Forward* economic and social reform policy from 1958 to 1962.[3] Such an horrendous disaster could have easily been prevented, by testing these policies in a province or island first, instead of forcing the economy as a whole into it. Even if the pilot project failed in the tested region, the rest of China could have helped feed them and learn from the gathered data and mistakes.

---

[*] Address and affiliation: University of Primorska, Faculty of Management, Koper; CM's email: 68223065[at]student.upr.si.
[1] http://web.archive.org/web/20101204064823/http://www.hsbc.com/1/2/about/history/1946-1979
[2] Matthew Evangelista, Peace Studies: Critical Concepts in Political Science, 2005, 96.
[3] frankdikotter.com/books/maos-great-famine/synopsis.html

On the other hand, Mao also killed more wilfully between 2 million and 5 million people in the 1950 to 1952 land reform,[4] who were mostly land owners or intellectuals –labelled 'capitalist', 'feudal' or 'counter-revolutionary'– or simply chosen to meet Mao's execution quotes. And also during Mao's *Cultural Revolution* from 1966 until his death in 1976 killed between 750 thousand and 1.5 million humans[5], again exclusively people in China and mostly intellectuals and everyone else perceived or simply declared as 'capitalist', 'feudal' or 'counter-revolutionary'.

This exemplifies how cruel, but also how extreme, restive and hostile with often deadly intent The People's Republic of China, the Communist Party of China and in particular Mao Zedong are: be it early on, right from 1949 onwards – or towards Mao's death in 1976. It is therefore incredible, how HSBC still managed to keep its Shanghai branch, potentially in the shadows, away from legal and vigilante enforcement.

Despite the still running office in Shanghai, due to the prevailing communism and anti-capitalism in the People's Republic of China, the opportunities for HSBC there were likely limited. That is, at least until Deng Xiaoping's reform of economic liberalisation gained momentum at the end of the 1980s and the early 1990s. Hong Kong –on the other hand– was booming economically, also for HSBC. Hong Kong remained HSBC's core operation, and to avoid a too concentrated –and not enough diversified– business model, HSBC expanded to and also focused more on other regions.[6] On 11 July 1955, the bank upgraded its San Francisco branch to an incorporated subsidiary, *(The) Hongkong and Shanghai Banking Corporation of California*, later more simply: *(The) Hongkong Bank of California*.[7,8,9] The bank also expanded by acquisitions, such as by fully buying in 1959 the Mercantile Bank –with its main operations in the Indian subcontinent and strong presence in Malaysia–, as well as the The British Bank of the Middle East in the same year, which was initially focused on Iran, yet left the Iranian market and focused on other regions in the Middle East instead.[10,11] Another acquisition, though what did not help much with the task of diversifying away from Hong Kong, was a majority stake in the Hong Kong based *Hang Seng Bank* in 1965.[12,13]

More expansions followed in the 1980s: The United States based *Marine Midland Bank* was added into the bank's holdings; first via a 51 per cent equity stake in 1980, before acquiring also the remaining 49 per cent in 1987.[14] As the 1955 incorporated Hongkong and Shanghai Banking Corporation of California was a subsidiary with primary presence on the West Coast, the acquisition

---

4   Lee Feigon, *Mao: A Reinterpretation*, Chicago 2002, 96.
5   Roderick MacFarquhar and Michael Schoenhals, *Mao's Last Revolution*, Cambridge (Massachusetts) 2006, 262.
6   http://web.archive.org/web/20101204064823/http://www.hsbc.com/1/2/about/history/1946-1979
7   history.hsbc.com/collections/global-archives/the-hongkong-and-shanghai-banking-corporation-of-california
8   http://web.archive.org/web/20101204064823/http://www.hsbc.com/1/2/about/history/1946-1979
9   history.hsbc.com/collections/global-archives/the-hongkong-and-shanghai-banking-corporation-of-california/records-relating-to-shares-and-shareholders/2009373-share-certificate-issued-by-the-hongkong-and-shanghai-banking-corporation-of-california
10  http://web.archive.org/web/20101204064823/http://www.hsbc.com/1/2/about/history/1946-1979
11  Geoffrey Jones, *The History of the British Bank of the Middle East*, Cambrdige (England) 1986.
12  http://web.archive.org/web/20101204064823/http://www.hsbc.com/1/2/about/history/1946-1979
13  www.grandprix.com/gpe/spon-022.html

of Marine Midland Bank with headquarters in New York helped expand to the East Cost of the United States.[15]

Yet this takeover is not to be confused with the takeover of a company that is similarly, just a word shorter named: The before mentioned London behemoth *Midland Bank*.[16] While the legal formality may have been a takeover, economically it was much closer to a merger of equal giants, resulting in two companies that each almost doubled in size post-merger, measured by assets.[17] Also, Midland Bank's roots go back further than those of HSBC, as the former date back to 22 August 1836: thereby somewhat extending HSBC's history by 29 years.[18]

The takeover of a bank so well established in the United Kingdom was in line with HSBC's strategic move away from Hong Kong and towards London: On 25 March 1991 –that is, in the year preceding the takeover of Midland Bank– HSBC changed its name to the as of 2025 current *HSBC Holdings plc* and its place of incorporation to London.[19] The additional listing on London –which resulted in a dual listing together with Hong Kong– allowed HSBC the funds to finance such a humongous takeover.[20] The year after the Midland Bank takeover, it moved its headquarters to London as well.[21]

Some observers believe the violent response by Deng Xiaoping and the Communist Party of China to the 1989 pro-democracy protests on Beijing's Tiananmen Square was the reason for the move away from Hong Kong[22]: On 4 June 1989, *People's Liberation Army* soldiers and their tanks killed hundreds and possible thousands of peaceful citizens exercising their constitutional right to demonstrate[23] – most of them young students, many on a hunger strike to be heard in their fight for a better, more just society.[24,25]

The Communist Party of China strikes this event from the history it teaches and attempts to censor it from the internet and everywhere else it can.[26] Yet this might have discouraged investors, including HSBC's management. And history might have proven them right:

By 2019, in then democratic Hong Kong, the annual remembrance commemoration of the 1989 Tienanmen massacre has become a remindful tradition on the importance of a liberal-democratic system. In 2019, the 30 year commemoration of the massacre in Hong Kong however –paired with protests against proposed law that would enable extradition of Hong Kong residents to the People's Republic of China– caused such wide civic activism and wide-scale protests by the Hong Kong population, that the Communist Party of China and its supporters introduced these and even more laws in Hong Kong effectively abolishing democracy, the independent press and many other civic liberties.[27] These laws were introduced when a global pandemic wreak havoc in the world, after a new coronavirus emerged near the Wuhan *Institute of Virology*, also in the year 2019 – which, conveniently for Xi Jinping and the Communist Party of China, quelled the until early 2020 incredibly strong protests, as it created fear and made it illegal for people in Hong Kong to assemble.[28,29,30,31]

---

14 http://web.archive.org/web/20101204185155/http://www.hsbc.com/1/2/!ut/p/kcxml/
04_Sj9SPykssy0xPLMnMz0vM0Y_QjzKLN4o39DIASYGYxqb6kShCBvGOCJEgfW99X4_83FT9AP2C3NCIck
dHRQCsuaq3/delta/base64xml/L3dJdyEvd0ZNQUFzQUMvNElVRS82XzJfMUlT
15 history.hsbc.com/collections/global-archives/hsbc-usa-inc-marine-midland-corporation
16 web.archive.org/web/20090713/hsbc.com/1/2/about/history/1980-1999
17 web.archive.org/web/20090713/hsbc.com/1/2/about/history/1980-1999
18 about.hsbc.co.uk/history-timeline
19 sec.gov/Archives/edgar/data/1089113/000102123106000157/b822899ex1-1.htm
20 www.grandprix.com/gpe/spon-022.html
21 qz.com/616791/hong-kong-has-probably-lost-hsbcs-headquarters-for-good-and-beijing-is-to-blame
22 qz.com/616791/hong-kong-has-probably-lost-hsbcs-headquarters-for-good-and-beijing-is-to-blame
23 english.gov.cn/archive/lawsregulations/201911/20/content_WS5ed8856ec6d0b3f0e9499913.html -> Article 35
24 https://web.archive.org/web/20190330013154/http://content.time.com/time/magazine/article/
0,9171,970278,00.html
25 Timothy Brook, *Quelling the People: The Military Suppression of the Beijing Democracy Movement,* Stanford (California) 1998, 154.
26 theage.com.au/world/china-tightens-information-controls-for-tiananmen-anniversary-20090604-bvxf.html

An extraordinarily interesting fact: Also in Hong Kong, also in 2019 –on 21 July– at the Yuen Long rail station, between 600 and 700 rod-wielding, suspected triad members beat up dozens and potentially hundreds of protesters when they returning home – all while police was idling nearby, arriving 39 minutes after the attack started and 1 minute after the attackers left, despite tens of thousands of emergency calls.[32,33,34,35,36,37,38] Less than six weeks later, Hong Kong police themselves repeated a similar, indiscriminate assault on civilian protesters returning home, also at one of Hong Kong's rail stations.[39]

Why is all that so interesting: Not only did HSBC help finance the rail network where these assaults took place, it also facilitated the opium trade which allowed triads to be a problem there until as of this writing in 2025[40,41], and finally, it's management also foresaw the eroding of the Hong Kong liberal-democratic system, as exemplified by the above events, long before such eroding occurred. This prediction can not be taken for granted: While the end of the 99 year lease was predictable, that lease only referred to the New Territories, not Hong Kong Island nor Kowloon, which legally – yet due to *unequal treaties*–, Britain could keep permanently. It was the British government who decided against a partial and in favour of a full handover. Further, the Communist Party of China promised Hong Kong to stay a liberal democracy at least until 1 July 2047.[42] Therefore, perhaps it was more risk aversion than prediction, paired with global and geopolitical understanding and an appreciation for optionality, that resulted in the move away from Hong Kong.

---

27  apnews.com/article/china-hong-kong-beijing-democracy-national-security-9e3c405923c24b6889c1bcf171f6def4
28  apnews.com/article/china-hong-kong-beijing-democracy-national-security-9e3c405923c24b6889c1bcf171f6def4
29  edition.cnn.com/2023/02/26/politics/covid-lab-leak-wuhan-china-intelligence/index.html
30  dni.gov/files/ODNI/documents/assessments/Report-on-Potential-Links-Between-the-Wuhan-Institute-of-Virology-and-the-Origins-of-COVID-19-20230623.pdf
31  bbc.com/news/world-us-canada-64806903
32  https://web.archive.org/web/20190726194102/https://www.washingtonpost.com/world/hong-kong-protesters-occupy-airport-taking-message-to-visitors/2019/07/26/f4b2ea62-af6b-11e9-9411-a608f9d0c2d3_story.html
33  https://web.archive.org/web/20190915073217/https://theinitium.com/article/20190723-hongkong-yuenlong-incident-timeline/
34  https://web.archive.org/web/20190722043736/https://news.rthk.hk/rthk/ch/component/k2/1469670-20190722.htm
35  https://web.archive.org/web/20190721220857/https://thestandnews.com/politics/%E5%85%83%E6%9C%97%E7%AB%99%E6%83%A1%E7%85%9E%E6%89%8B%E6%8C%81%E6%9C%A8%E6%A3%92%E6%89%93%E4%BA%BA-%E5%B8%82%E6%B0%91%E8%A8%98%E8%80%85%E8%A2%AB%E8%BF%BD%E6%89%93%E5%8F%97%E5%82%B7-%E6%9C%AA%E8%A6%8B%E8%AD%A6%E5%93%A1%E5%9F%B7%E6%B3%95/
36  https://web.archive.org/web/20190722221837/https://www.scmp.com/news/hong-kong/law-and-crime/article/3019524/least-10-injured-baton-wielding-mob-suspected-triad
37  https://web.archive.org/web/20190724015415/https://news.mingpao.com/ins/%E6%B8%AF%E8%81%9E/article/20190723/s00001/1563890883621/%E3%80%90%E5%85%83%E6%9C%97%E8%A5%B2%E6%93%8A%E3%80%91%E7%99%BD%E8%A3%99%E5%A5%B3%E6%87%B7%E5%AD%95%E4%B8%8D%E8%B6%B3%E5%80%8B%E6%9C%88%E6%B2%92%E9%80%9A%E7%9F%A5%E9%86%AB%E9%99%A2-%E8%A8%BA%E6%89%80%E6%B1%82%E9%86%AB%E8%AD%89%E8%83%8E%E5%B9%B3%E5%AE%89
38  youtube.com/watch?v=sicZ5TRdO2A
39  upload.wikimedia.org/wikipedia/commons/9/98/HK_police_storm_Prince_Edward_station_and_attack_civilians_20190831_11pm.webm
40  https://arteptweb-a.akamaihd.net/am/ptweb/117000/117100/117180-001-A_SQ_0_VA-STA_09586370_MP4-2200_AMM-PTWEB-80921910227504_2UENk7rOfO.mp4
41  https://arteptweb-a.akamaihd.net/am/ptweb/117000/117100/117180-002-A_SQ_0_VA-STA_09586380_MP4-2200_AMM-PTWEB-80921910227723_2UENl7rOfO.mp4

## HSBC 1990s to 2025: Post cold war, crises and financial results

HSBC's current hexagon logo was adopted in November 1998. HSBC continued to acquire even more and more other banks since the 1990s: In Argentina and Brazil in 1997 and in 1999, the Republic New York Corporation. In 2000, HSBC acquired Crédit Commercial de France, in 2001 a real estate company in Australia, Demirbank in Turkey, a 97 per cent of a Taiwan's largest asset management company and so on and so forth – In short: HSBC grew organically and through acquisitions.[43]

Yet, the Global Financial Crisis of 2005-2010 also affected HSBC: It could stay afloat without government assistance, though it had to ask for cash from its shareholders in HSBC's 17.7 Billion US dollar Right Issue cast in 2009.[44] In 2010, it focuses more on Hong Kong and the People's Republic of China, as well as on the fast developing jurisdictions of Vietnam and Sri Lanka.[45]

The following table and graph better visualise the growth, as well as the ups and downs and other metrics of interest in regard to HSBC from the late 1980s until its latest report year as of 2025: 2024.

---

42  web.archive.org/web/20160620182610/http://www.nytimes.com/learning/general/onthisday/big/0630.html
43  web.archive.org/web/20100728014233/http://www.hsbc.com/1/2/about/history
44  http://web.archive.org/web/20100728014233/http://www.hsbc.com/1/2/about/history
45  http://web.archive.org/web/20100728014233/http://www.hsbc.com/1/2/about/history

**Table 1:** HSBC's assets and profits, by year, in million (mio) United States dollars (USD).

| Year | Assets | Profit |
|---|---|---|
| 1987 | 105,770 | 461 |
| 1988 | 112,121 | 1,009 |
| 1989 | 136,409 | 1,015 |
| 1990 | 149,663 | 565 |
| 1991 | 155,187 | 1,105 |
| 1992 | 283,729 | 1,574 |
| 1993 | 309,573 | 2,695 |
| 1994 | 314,771 | 3,149 |
| 1995 | 352,022 | 3,885 |
| 1996 | 402,377 | 4,852 |
| 1997 | 471,686 | 5,487 |
| 1998 | 483,128 | 4,318 |
| 1999 | 569,908 | 5,408 |
| 2000 | 673,503 | 6,457 |
| 2001 | 695,545 | 4,992 |
| 2002 | 758,605 | 6,239 |
| 2003 | 1,034,216 | 8,774 |
| 2004 | 1,279,974 | 12,918 |
| 2005 | 1,501,970 | 15,060 |
| 2006 | 1,860,758 | 15,699 |
| 2007 | 2,354,266 | 19,043 |
| 2008 | 2,527,465 | 5,546 |
| 2009 | 2,364,452 | 5,565 |
| 2010 | 2,454,689 | 12,746 |
| 2011 | 2,555,579 | 16,224 |
| 2012 | 2,692,538 | 13,454 |
| 2013 | 2,671,318 | 15,631 |
| 2014 | 2,634,139 | 13,115 |
| 2015 | 2,409,656 | 12,572 |
| 2016 | 2,374,986 | 1,299 |
| 2017 | 2,521,771 | 9,683 |
| 2018 | 2,558,124 | 12,608 |
| 2019 | 2,715,152 | 5,969 |
| 2020 | 2,984,164 | 3,898 |
| 2021 | 2,957,939 | 12,607 |
| 2022 | 2,949,286 | 14,346 |
| 2023 | 3,038,677 | 22,432 |
| 2024 | 3,017,048 | 22,917 |

Profit = Profit attributable to ordinary/common shareholders of the parent company (HSBC).

*Sources:*

— https://fred.stlouisfed.org/data/EXHKUS

— https://fred.stlouisfed.org/data/EXUSUK

— https://www.sec.gov/cgi-bin/browse-edgar?action=getcompany&CIK=0001089113&type=20&dateb=&owner=include&count=100&search_text=

— https://web.archive.org/web/20251106135752if_/https://www.udrop.com/O0vo/2025-01-21HsbcAr1991-2023ex2010.zip?download_token=c02d44d874887d5a8315a297a13e46ddea6b3a8cf26feb18efa9dd4d1c5ea4c6

**Figure 1:** HSBC's assets in Trillion (Tr) and profits in Billion (Bn) USD, by year.

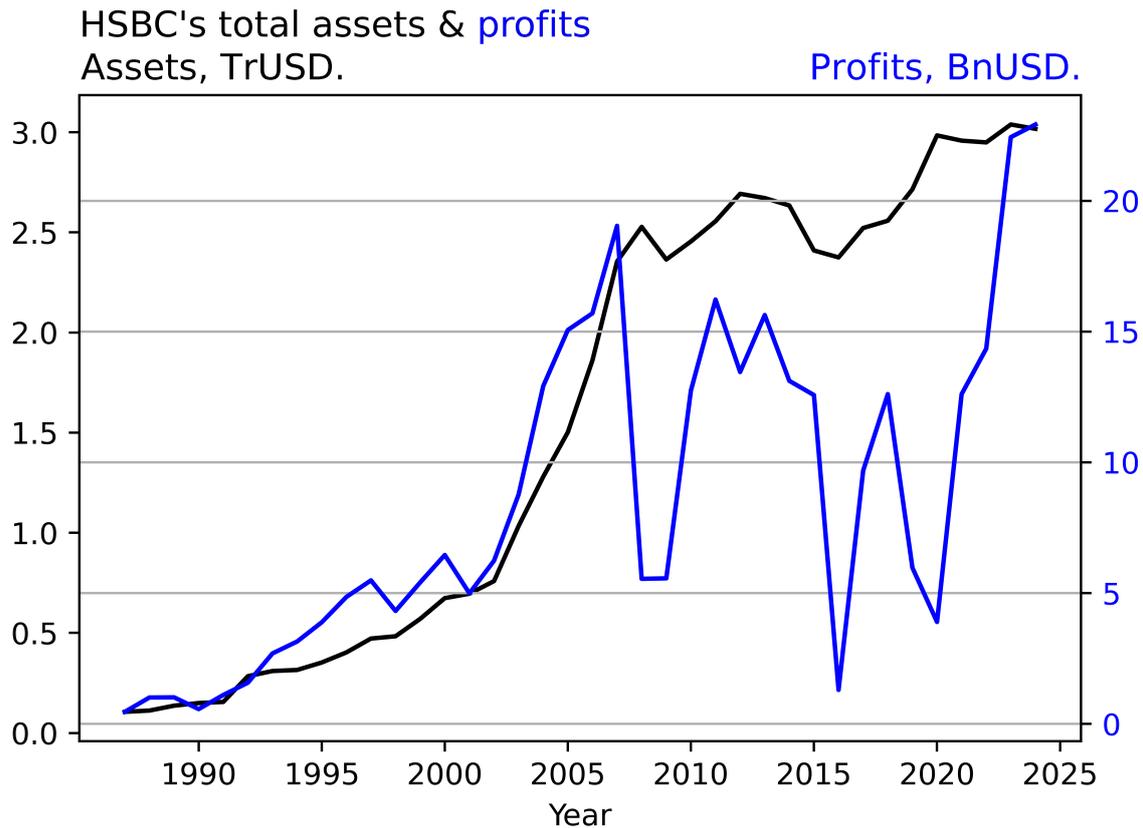

Sources: See under Table 1. See also spreadsheet used to compile this data at: perma.cc/PHP7-G4CN

In the observed period, the following crises occurred:
— Black Monday of 1987
— Black Wednesday of 1992
— 1997-1998 Asian financial crisis
— 1998-2003 Dotcom boom and bust
— 2005-2010 Global financial crises
— 2020-2022 Global pandemic

Yet, what is astonishing to see: HSBC claims to have remained profitable in every single year between 1987 and 2024. Even during the Great Recession around 2008, when many banks went under or had to be rescued by the government and taxpayers. This is also when HSBC asked for cash from its shareholders[46]. Yet still, the company declared profits for every year from 1987 to 2024 – and not a single annual loss.

---

46  http://web.archive.org/web/20100728014233/http://www.hsbc.com/1/2/about/history

# List of possible reasons for survival and success

So, in light of all of the above, how could any company, let alone a financial institution –which arguable needs stability more than other industries and is more prone to suffer from drastic changes or chaos– become so successful, in spite of the astonishingly turbulent history of the place where it was incorporated, headquartered and is main place of business was, for most of its history?

There are a number of possible reasons:
— Preceding its founding, business in Hong Kong was booming. 70 per cent of it might have been illegal and incredibly harmful opium trade, yet that was irrelevant for HSBC, its good business and profits.
— That leads to the next possible reason: Business and profit comes first, moral values or what is best for society or one's surrounding fellow human beings comes at HSBC –at best– second. That was the case in 1865 as it is in 2025: Countless controversies run like a thread through the banks history, starting with colonial injustices after subjugation through war, which enabled it to be founded. Continuing through the illegal and amoral opium trade and the triad societies it thereby helped, which plague the Mandarin and Cantonese speaking communities globally until this day as of this writing in 2025.[47,48,49,50,51,52] It shows in the human resource system, which for most of its time, discriminated against some of its staff by attempting to divide the one human race into multiple groups, "into three ethnic tiers" – such practices are commonly called racism, which assumes what has been disproved by biological and medicinal research.[53,54] While HSBC abandoned the system in place officially in the 1960s[55], there are reports at least until 2021 that discrimination is still a major issue at the bank.[56,57] In addition to all of that, an incredible amount of controversies and plausible allegations seem to continuously plague the bank, some of which are money laundering, facilitating transactions of drug trafficking proceeds, tax evasion, market manipulation and fraud, including defrauding its own clients.[58,59,60,61]

This all goes in addition to minor, yet quite telling unethical practices: As pointed out further above, HSBC does not mention the colonial background that allowed its foundation, nor its opium heavy

---

history when it claims to inform about its past.[62,63,64,65,66,67,68,69,70,71,72] Also, HSBC foresaw –or at least saw the risk of– the illegal attacks on the liberal-democratic system in Hong Kong, committed by the Communist Party of China[73] – indicated by its move away from Hong Kong to London in 1991 and 1992. It did not use its major weight in Hong Kong –where it is often merely just referred to as 'The Bank' and issues between 40 to 50 per cent of its legal tender notes[74,75]– to try to protect its democracy; HSBC seemingly felt no responsibility for the place where it grew to become what it is today.

— When discussing the early years of HSBC, not only the colonial and economic environment contributed to HSBC's success story, but also the conditions that Thomas Sutherland set: The well-educated and well-connected Sutherland started HSBC in a centrally located, large, brand new, high-quality building, equipped it with a lot of starting capital, conservative principles and proper articles of association and also helped it establish itself from day one through the connections he already had, and kept building, at P & O. Thomas Jackson –when handed the baton after Sutherland– continued that momentum and grew the bank even more than under Sutherland. This solid foundation likely tremendously increased its chances of surviving the geopolitical disasters of the following civil and world wars. As the saying goes, *one must make hay when the sun shines* – and prepare for the lean years.

— HSBC continued to display a crucial element in risk minimisation – made popular by the investment paradigm with the same name: *Diversification*. HSBC diversified early on and never stopped. It diversifies its chances and its risks. It has always been diversifying in terms of regions – as shown early on with the Shanghai office and even evident in the bank's names.[76] The bank also is diversified in terms of segments, which are very balanced as of 2023 – albeit all segments being necessarily within the financial industry.[77] One aspect of diversification is the *diversification in assets in the narrower sense* – such as *different kinds of assets* (physical stocks, bonds, precious medals etc), and them being stored *at different places*. Another aspect is the diversification of *assets in the wider sense*, such as having human resources, client and other business networks and independent revenue streams in multiple places. Another crucial aspect of regional diversification – paired with the flexibility of the globalised world of finance– allowed HSBC to simply shift its place of business, even of its place of incorporation and its headquarters, away from potential risks. It did so during World War II, when the bank temporarily moved to London before returning to Hong Kong; or before the handover of Hong Kong to the Communist Party of China and the

---

eroding of the liberal-democratic and rule of law based system in Hong Kong that followed, when the bank moved more permanently to London in 1991 and 1992, where the bank has its headquarters and place of incorporation since. If a crisis where to emerge in the United Kingdom, HSBC could –and likely would– move both to Paris, New York City, Amsterdam, Dublin, Sydney, Frankfurt or any other fitting city or jurisdiction. The case study of HSBC shows, always picking the most advantageous places anywhere in the world for any period can be incredibly beneficial. Though, only few have that option. Equipment heavy or location dependent industries, such as factories or construction companies may lack that precious possibility. Nevertheless it could be this independence of and diversification in regions, paired with further diversification of assets, risks etc and the creation if independent subunits that allowed the bank to avoid a terminal, finite, total loss. Or –as shown above– even any loss for any year, at least between 1987 and 2024.

— Lastly and maybe the least intuitive point, especially compared to the above: The very reason of HSBC's resilience might lie in the chaotic environment itself – aside from an obvious selection bias, as this case study would not exist if HSBC failed. HSBC was forced from its very first day to be in constant friction with the law and governments, with organised crime and with geopolitical disasters. It was forced to survive in such an environment, which makes it aware of, understand, adapt to and prepare for all these risks – as well as comprehend the chaotic, unforeseeable, extreme nature of our world and act accordingly. On the other hand, no major war nor serious threat to its liberal-democratic system occurred in the United States since the end of the United States Civil War, which ended in 1865. Yet even in the past 20 years, there are hundreds of bank failures, including some banks who failed with hundreds of billions of US dollars in assets.[78] And even more bankruptcies would have been practically guaranteed, if US federal institutions had not prevented them with public and tax payer money. Hong Kong, on the other hand, had zero bank failures since its last one in July 1991 – Even though the last bank to collapse caused a bank run and distrust in other banks, which Hong Kong's other banks weathered.[79,80] Not even the Black Wednesday of 1992 –exacerbated by Hong Kong latest bank failure as of now, in just the year before–, the 1997-1998 Asian financial crisis –which was much closer to Hong Kong than to the US–, the 1998-2003 Dotcom boom and bust, the 2005-2010 Global Financial Crises and Recession, the 2015–2016 Chinese stock market turbulence –again: much closer intertwined with Hong Kong than the US– nor the 2020-2022 Global Pandemic could bring down any of Hong Kong's resilient banks. Even when adjusting for population –there were 569 bank failures in the US between 2001 and early 2025 and Hong Kong has 2.2 per cent the population of the US, hence– it would result in an expected >12.5 bank failures in Hong Kong in the same period – when applying the same bank bankruptcies per population basis. Yet, there were none.

This resilience through stressors concept –with wide applications, including business, social and government structures, evolution, sports: training muscles, psychology or immunology– was illustrated and popularised by Prof Dr Nassim Nicholas Taleb, including in his 2012 book "Antifragile".[81]

---

78  fdic.gov/resources/resolutions/bank-failures/in-brief/index.html
79  ft.com/content/154e4c66-ee65-11e4-98f9-00144feab7de
80  dps.org.hk/en/background.html
81  https://search.worldcat.org/title/1337501406

# Conclusions

To summarily conclude what entrepreneurs can learn from HSBC's story: The case study of HSBC shows that the following reasons may contribute to long-term survival and financial success – Many of which might not be surprising, yet could be essential for success nonetheless:

─ Ethics is not essential for prosperity – making money is. It is unknown whether that is more true for larger, global companies than for smaller, local ones, which might be closer connected to the locality and its individuals.
─ The start is important. Founders should ideally start at a good place, in well-situated area, with a large network of business partners and with plenty of starting capital. They should lay down the right principles and articles of association for the company's long term success.
─ Diversify in every way: Regionally and in segments, horizontally and vertically. The more independent one's units are from each another, the lower the chance that all will fail at the same time. Locations, assets, human resources, business partners, revenue streams and the dependence, independence or interdependence of all of them should all be part of the thought process when thinking about risk minimisation through diversification, to maximise long term survival.
─ Gathering information and growing through stressors. Just like one trains one's muscle with stress, companies which experienced adverse conditions may be better adapted to future adverse scenarios. Entrepreneurs should ensure they learn from all information, including –and maybe especially– in regard to unfavourable conditions.